*Topical review for SPQEO*

# NANOFERROICS: STATE-OF-ART, GRADIENT-DRIVEN COUPLINGS AND ADVANCED APPLICATIONS (Authors' review)

**Anna N. Morozovska[1], Ivan. S. Vorotiahin[1,2], Yevhen M. Fomichov[3,4], and Cristian M. Scherbakov[5]**

[1]Institute of Physics, National Academy of Sciences of Ukraine,
46, Prospekt Nauky, 03028 Kyiv, Ukraine

[2] Institut für Materialwissenschaft, Technische Universität Darmstadt, Jovanka-Bontschits-Str. 2, 64287 Darmstadt, Germany

[3] Faculty of Mathematics and Physics, Charles University, V Holesovickach 2, 18000 Prague 8, Czech Republic

[4]Institute for Problems of Materials Science, National Academy of Sciences of Ukraine,
3, Krjijanovskogo, 03142 Kyiv, Ukraine

[5] Taras Shevchenko Kiev National University, Physical Faculty, Chair of Theoretical Physics,
4e, pr. Akademika Hlushkova, 03022 Kyiv, Ukraine

**Abstract**

Ferroics and multiferroics are unique objects for fundamental physical research of complex nonlinear processes and phenomena, which occur in them in micro and nano-scale. Due to the possibility of their physical properties control by size effects, nanostructured and nanosized ferroics are among the most promising for advanced applications in nanoelectronics, nanoelectromechanics, optoelectronics, nonlinear optics and information technologies. The review discuss and analyze that the thickness of the strained films, the size and shape of the ferroic and multiferroic nanoparticles are unique tools for controlling their phase diagrams, long-range order parameters, magnitude of susceptibility, magnetoelectric coupling and domain structure characteristics at fixed temperature. Significant influence of the flexo-chemical effect on the phase transition temperature, polar and dielectric properties is revealed for thin films and nanoparticles. Obtained results are important for understanding of the nonlinear physical processes in nanoferroics as well as for the advanced applications in nanoelectronics.

## 1. INTRODUCTION

**1.1. Definition of ferroics and multi ferroics.** Under the definition, "classical" ferroics are condensed substances which under the certain external conditions (with changing temperature, pressure, etc.) occurs a spontaneous symmetry lowering, as the result is the collective phase transition of its elementary structure into an ordered state [1, 2]. In this state, a vectorial (or tensorial) long-range order parameter arises, whose direction (or magnitude), as a rule, can be changed between several (meta)stable states by applying physical fields larger than the coercive value. The phenomenon is named ferro-hysteresis. Some ferroics in the ordered state are spontaneously divided into "domains" – micro- or nano-scopic regions with a certain direction (or magnitude) of the order parameter. "Classical" first-order ferroics are solid-state ferromagnets, ferroelectrics, antiferroelectrics, ferroelastics, substances with antiferrodistorsion. Magnetic and ferroelectric relaxors, quantum paraelectrics, superparamagnetics, superparaelectrics, ferroelectric liquid crystals, ferrionics sometimes can be regarded ferroics. This is due to the fact that, in accordance with the creator of the name "ferroic" Aizu [3, 4] the possibility of a phase transition induced by an external field near some critical temperature $T_C$ is a general feature of both ordered and disordered ferroics.

The long-range order parameter of ferromagnetics is the vector of the spontaneous magnetization of the lattice **M**(T,H), which occurs below the Curie temperature due to the ordering of the orientations of elementary spins [the phase transition of «order-disorder» type, see **Fig.1(a)**] and has a hysteretic behavior as a function of the external magnetic field **H** [see **Fig.1(b)**]. There is a ferromagnetic domain structure below $T_C$.

The long-range order parameter of antiferromagnetics is the vector of spontaneous magnetization of individual sublattices **L**(T,H), which occurs below the Néel temperature $T_N$. The magnetization of antiferromagnetics **M** has an "anti" hysteresis behavior as a function of the external magnetic field **H** [see **Fig.1(c)**].

The long-range order parameter of ferroelectrics is vector of spontaneous polarization of the lattice P(T,E), which occurs below the "ferroelectric Curie temperature " $T_{FE}$ due to the appearance or ordering of the orientation of elementary dipoles [phase transitions of the "bias" type or "ordering-disordering" type depending on concrete material, see **Fig.1(a)**] and has a hysteretic behavior as a function of the external electric field E [see **Fig.1(b)**]. There is a domain ferroelectric structure below $T_{FE}$.

The long-range order parameter of antiferroelectrics is vector of spontaneous polarization of individual sublattices **A**(T,E), which occurs below the temperature $T_{AFE}$. Polarization of antiferroelectrics has an "anti" hysteresis behavior as a function of the external electric field E [see **Fig.1(c)**]. The domain structure does not exist in uniaxial antiferroelectrics.

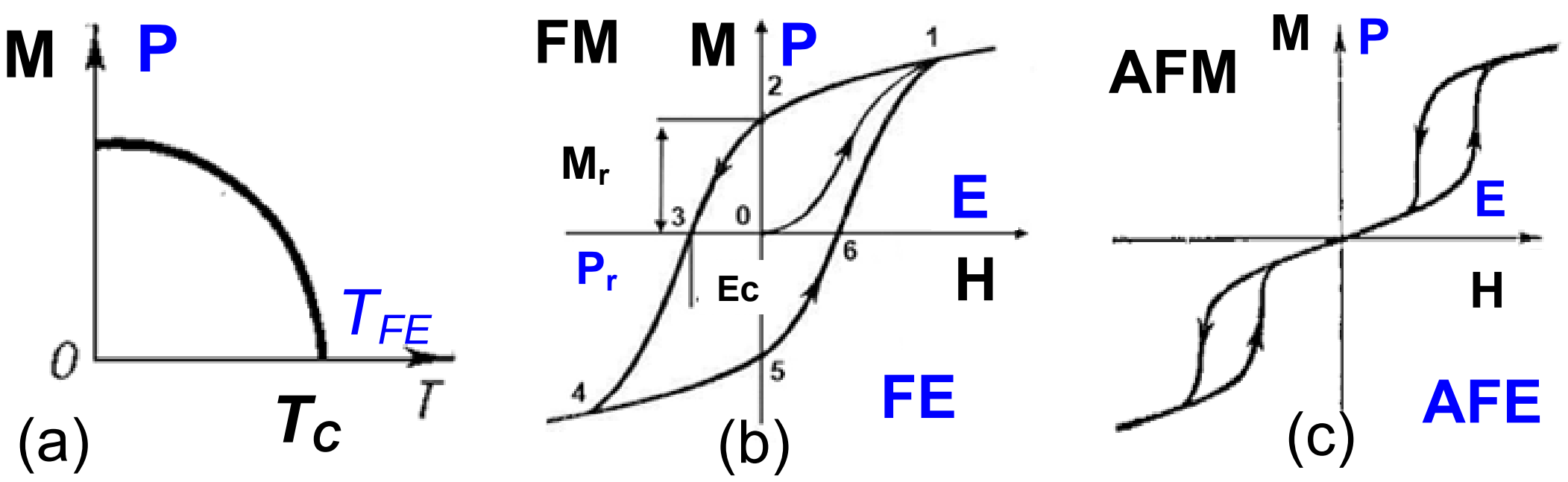

**Fig. 1. (a)** A typical temperature dependence of the order parameter in ferromagnetics and ferroelectrics. **(b)** Hysteresis behavior of the order parameter as a function of the external field below the temperature of the phase transition. **(c)** Double hysteresis ("antihysteresis") of the order parameter as a function of the external field below the temperature of the phase transition, distinctive for antiferromagnets and antiferroelectrics. Adapted from Ref.[[5]].

The long-range order parameter of ferroelastics is the component the spontaneous lattice strain $u(T,p)$, which has hysteretic behavior as a function of external pressure $p$ and arises below the temperature of the phase transition due to the spontaneous symmetry lowering in the elementary lattice. Sometimes the ferroelastic phase transition is accompanied by the appearance of the secondary order parameter – polarization (that is also registered experimentally).

The structural order parameter in ferroics with antiferrodistorsion is the pseudovector of the angle of static rotation of certain crystallographic groups $\Phi(T,p)$, which arises spontaneously below the temperature of the phase transition due to the spontaneous symmetry lowering (flexure) in the unit cell (see **Fig.2**). In this case the ferroic is broken down into «elastic» domains - "twins", which are registered experimentally. Static rotations of the crystallographic groups can be detected by scanning electron microscopy with an ultra high atomic and subatomic resolution [[6]]. It should be noted that Aizu for the first time gave the name of the definition of ferroelasticity as a feature, which can exist by itself in crystals and not the same with ferroelectrics and ferromagnetics [3-4].

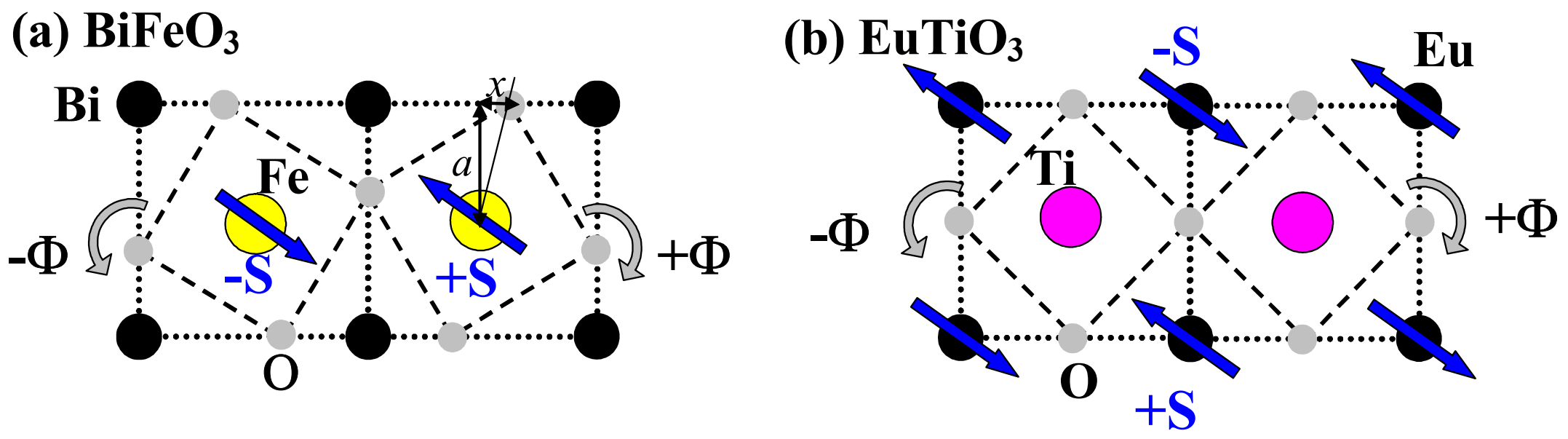

**Fig. 2.** The structural antiferrodistortive order parameter is a pseudovector of the angle of rotation of oxygen octahedra in multiferroics $BiFeO_3$ **(a)** and $EuTiO_3$ **(b)**. Adapted from Ref.[7]

Ferroics are consolidated by the same behavior of the main characteristics, such as magnetic, electrical and mechanical. Ferroics that have simultaneously more than one of these characteristics are called multiferroics. Thus, the difference between a ferroic and a multiferroic is that the multiferroic is a "complex" ferroic of the second (or higher) order in which two (or more) long-range order parameters of different physical nature coexist and interact under certain external conditions. In this case, the external field that induces hysteresis behavior of one order parameter will induce hysteresis behavior of another one due to the coupling between these parameters. The susceptibility and domain structure of different types are interrelated in multiferroics, which make them unique objects for fundamental and applied physical research [8].

**1.2. The influence of temperature and external fields on the physical characteristics of ferroics.** Temperature, external electric and magnetic fields, pressure (strain or deformation) are the main tools for controlling phase state, the magnitude of the susceptibility and the features of the domain structure in macroscopic ferroics and multiferroics. **Fig.3** shows the temperature dependence of the long-range order parameters in the multiferroic $BiFeO_3$, from which it can be seen that as the temperature decreases first appears spontaneous polarization in the material (below 1100 K) and then antiferromagnetic ordering (below 650 K). The antiferrodistortive order parameter arises in the vicinity of 1400 K, changes its symmetry (i.e. the type and number of the components) at 1100 K and 1300 K.

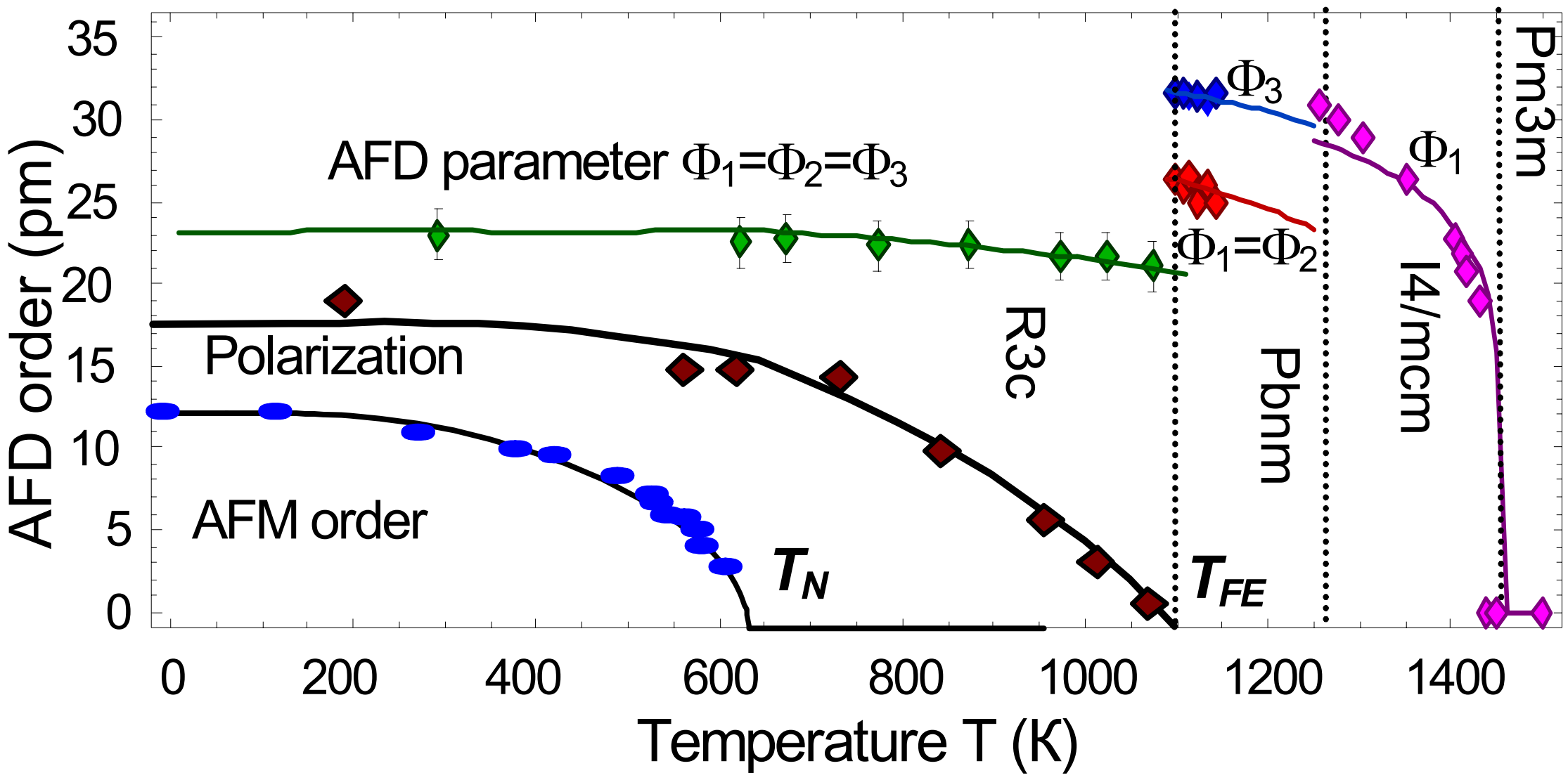

**Fig. 3.** Temperature dependence of order parameters in a multiferroic $BiFeO_3$. Vertical dashed lines denote transitions between different phases. The vertical designation denotes the designations of different symmetries (cubic Pm3m, tetragonal I4/mcm, orthorhombic Pbnm and rhombohedral R3c), which correspond to each phase. Adapted from Ref. [[9]].

Over the past 10 years, it has been established that the thickness of strained films, the size and shape of nanoparticles of ferroics and multiferroics are unique tools for controlling their phase diagrams, the values of order parameters and susceptibility, the magnetoelectric coupling, and the peculiarities of the domain structure at a fixed temperature [7, 9, [10]]. For example, **Fig. 4(a)** shows the dependence of the main components of the magnetoelectric coupling tensor on the radius of the multiferroic nanorod. It can be seen from the figure that the magnetoelectric coupling can increase in hundreds and thousands times with decreasing radius.

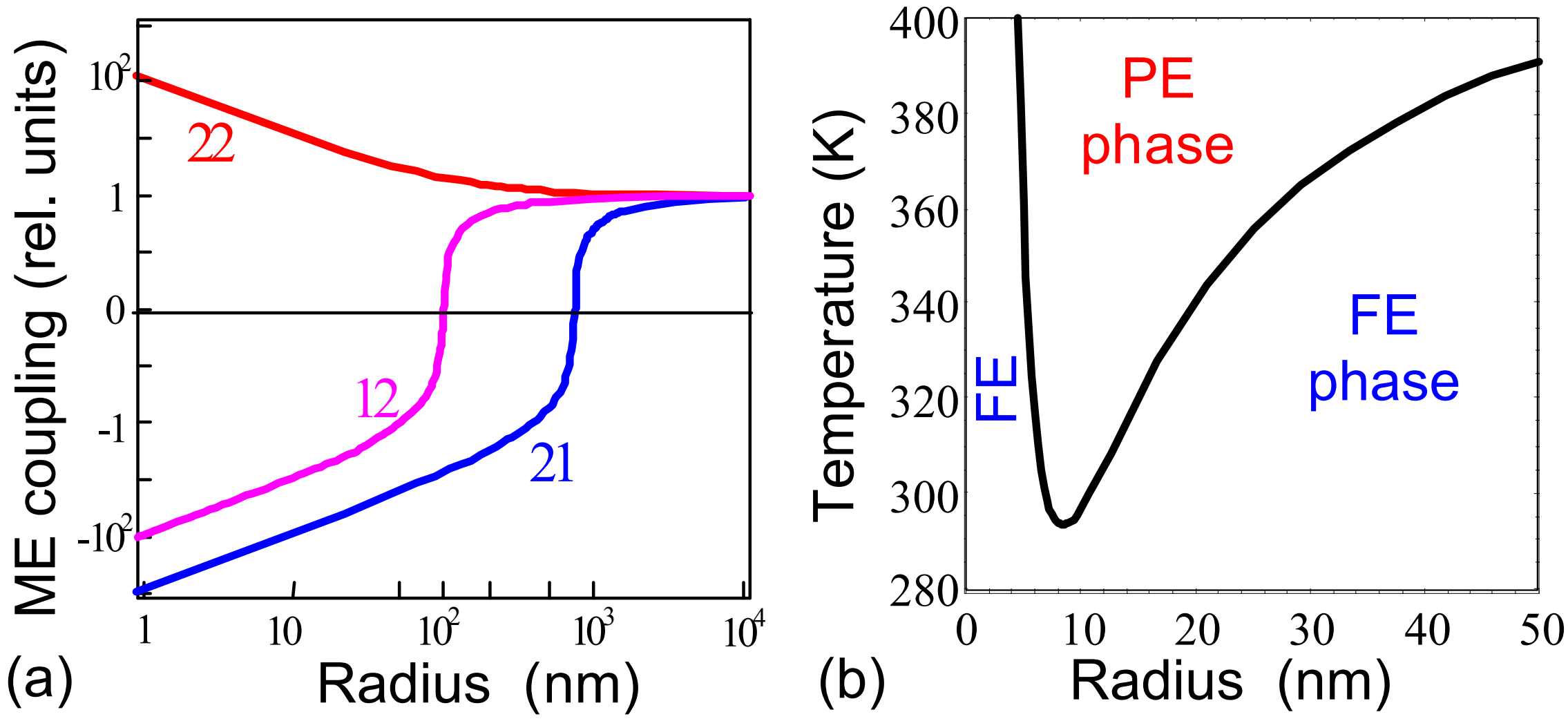

**Fig. 4. (a)** Dependence of the main components of the biquadratic magnetoelectric coupling tensor (12, 21, 22) on the radius of multiferroic nanoparticle $BiFeO_3$. Adapted from Refs. [5, 10]. **(b)** Phase diagram of a nanoparticle $BaTiO_3$ in the coordinates "temperature-particle radius". PE – non-polar paraelectric phase, FE - polar ferroelectric phase. Adapted from Refs. [5, 11].

### 1.3. Influence of the flexochemical effect on the physical characteristics of ferroics.

Recently, we observed a significant effect of the flexochemical coupling on the temperature of the phase transition, the polar and dielectric features of nanoferroics [[11], [12]], because there is spontaneous flexoeffect in the nanostructures revealed by Glinchuk et al. [[13]].

The flexoelectric effect is a linear relationship between the strain gradient and electric polarization (direct effect); and between the gradient of polarization and the strain (inverse effect). [[14]]. Chemical pressure is the mechanical stresses arising in a material as a result of changes in the dimensions of the crystal lattice around elastic defects [[15]]. For example, **Fig.4(b)** shows the phase diagram of a nanoparticle $BaTiO_3$ in the coordinates "temperature-particle radius". It is seen from the diagram that the nonpolar paraelectric phase (PE), which is usually stable at high temperatures and small particle radii, again turns into the polar ferroelectric phase (FE) in accordance with experiment of Zhu et al [[16]]. This is due to the size effect of flexochemical coupling.

It was found that the flexochemical coupling can induce the shear surface acoustic waves in non-piezoelectric materials [[17]], in this case, the depth of penetration of a such wave is determined not only by the magnitude of the static flexoelectric coupling (coupling constant $f$), but also by the value of the dynamic flexo-effect, that is, a polarization reaction to the accelerated motion of the medium (coupling constant $M$). There are no surface shear waves in matter without an inversion center under the condition $f=M=0$. When $f$ is greater than the critical value, a spatially modulated phase occurs (see. **Fig.5**).

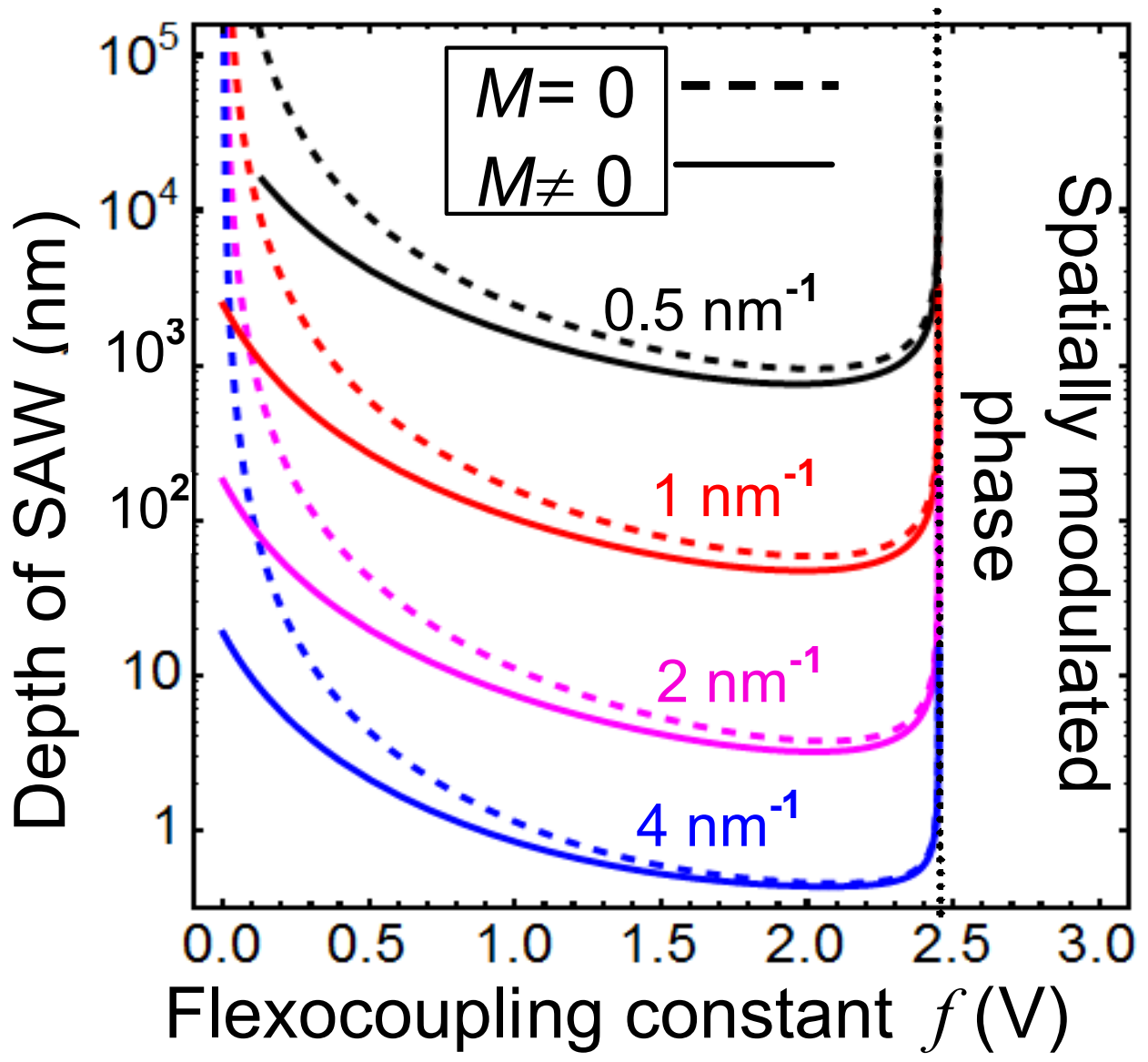

**Fig. 5.** Dependence of the penetration depth of a surface acoustic wave (SAW) in a paraelectric $SrTiO_3$, induced by a static and dynamic flexoelectric effects. Each pair of curves corresponds to the corresponding values of the wave vectors, whose magnitude is shown near the curves in reverse nanometers. Adapted from Ref. [17].

Flexoelectric coupling can induce a soft acoustic mode and incommensurate spatially modulated phases in ferroelectrics [[18], [19], [20]]. At the same time, a soft acoustic mode appears with increasing the flexoelectric coupling constant above the critical value, and there is a gap in the dispersion curve of the soft acoustic mode. A spatially modulated phase appears simultaneously with the gap (see **Fig. 6**). This fact makes it possible to understand the behavior and quantitatively describe the dielectric susceptibility and the spectra of soft phonons in different ferroelectrics with incommensurate phases, for which the existence and type of the soft mode remained unknown [20]. The result can be important for physical understanding of the nature of soft acoustic modes and the appearance of incommensurate phases in ferroelectrics and can be used to predict their phase diagrams, polar and electromechanical features, as well as to determine the coupling constant from phonon spectra.

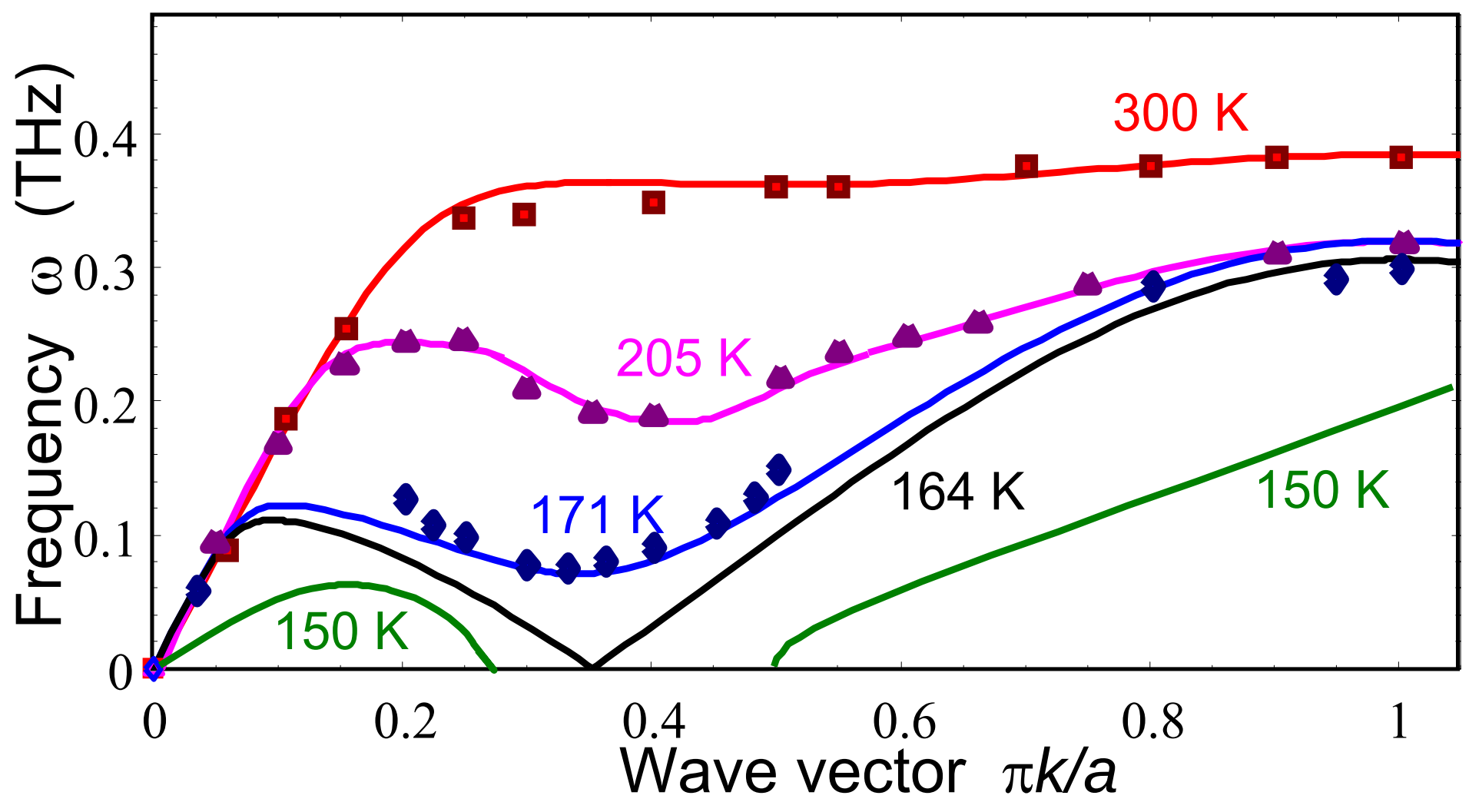

**Fig. 6.** The dependence of an acoustic mode frequency on its wave vector (in the reciprocal lattice constant units π/a). Symbols are experimental data from [21], measured in the organic ferroelectric $(CH_3)_3NCH_2COO{\cdot}CaCl_2{\cdot}2H_2O$ at temperatures from 300 K to 171 K. Solid curves are calculated at temperatures from 300 K to 150 K. Adapted from Ref. [20].

## II. FLEXO-CHEMICAL EFFECT IN THIN FILMS

### 2.1. Theoretical formalism

The Landau-Ginzburg-Devonshire (LGD) Gibbs free energy of the ferroelectric film consists of bulk ($G_V$) and surface ($G_S$) parts and the energy of the electric field outside the film ($G_{ext}$),

$$G = G_V + G_S + G_{ext}, \tag{1a}$$

$G_V$ and $G_S$ are the power expansion on the polarization vector and stress tensor components $P_i$ and $\sigma_{ij}$ [22]:

$$G_V = \int_{V_{FE}} d^3r \left( \begin{array}{l} \frac{a_{ik}}{2}P_iP_k + \frac{a_{ijkl}}{4}P_iP_jP_kP_l + \frac{a_{ijklmn}}{6}P_iP_jP_kP_lP_mP_n + \frac{g_{ijkl}}{2}\left(\frac{\partial P_i}{\partial x_j}\frac{\partial P_k}{\partial x_l}\right) - P_iE_i \\ -\frac{\varepsilon_0\varepsilon_b}{2}E_iE_i - \frac{s_{ijkl}}{2}\sigma_{ij}\sigma_{kl} - Q_{ijkl}\sigma_{ij}P_k P_l - F_{ijkl}\left(\sigma_{ij}\frac{\partial P_l}{\partial x_k} - P_l\frac{\partial \sigma_{ij}}{\partial x_k}\right) - W_{ij}\sigma_{ij}\delta N \end{array} \right), \tag{1b}$$

$$G_S = \int_S d^2r \left( \frac{a_{ij}^S}{2}P_i\,P_j - \frac{\varepsilon_0}{2\lambda}\varphi^2 \right), \qquad G_{ext} = -\int_{\vec{r}\notin V_{FE}} d^3r \frac{\varepsilon_0\varepsilon_e}{2}E_iE_i\,. \tag{1c}$$

The tensor $a_{ij}$ is positively defined for linear dielectrics, and explicitly depends on temperature $T$ for ferroelectrics and paraelectrics, and $a_{ij} = \alpha_T(T - T_c)\delta_{ij}$, where $\delta_{ij}$ is the Kroneker delta symbol, $T$ is absolute temperature, $T_c$ is the Curie temperature. Other tensors included in the free energy (1) are supposed to be temperature independent. Tensor $a_{ijklmn}$ should be positively defined for the thermodynamic stability in paraelectrics and ferroelectrics. The gradient energy tensor $g_{ijkl}$ regarded positively defined. $\varepsilon_0$ is the vacuum permittivity, $\varepsilon_b$ is a relative background dielectric permittivity [23]. Coefficients $Q_{ijkl}$ are the components of electrostriction tensor; $s_{ijkl}$ are the components of elastic compliance tensor, $F_{ijkl}$ is the flexoelectric strain coupling tensor. For most of the cases one can neglect the polarization relaxation and omit high order elastic strain gradient terms if the flexoelectric coefficients are below the critical values $F_{ijkl}^{cr}$ [24, 25]. Following Refs. [26, 27] we assume that the used approximation is valid if $F_{klmn}^2 << g_{ijkl}s_{ijmn}$. $W_{ij}$ is the Vegard strain tensor, that is regarded diagonal hereinafter, i.e. $W_{ij} = W\delta_{ij}$. The quantity $\delta N = N(\vec{r}) - N_e$ is the difference between the concentration of defects $N(r)$ at the point **r** and their equilibrium (average) concentration $N_e$. The tensor coefficients $a_{ij}^S = \alpha_{S0}\delta_{ij}$. Here we introduce electric field via electrostatic potential $\varphi$ as $E_i = -\partial\varphi/\partial x_i$ and an effective surface screening length $\lambda$ that can be much smaller than lattice constant [28]. Electric field $E_i$ includes external and depolarization contributions (if any exists).

Note that we include only one half ($F_{ijkl}P_k(\partial\sigma_{ij}/\partial x_l)$) of the full Lifshitz invariant $F_{ijkl}\left(P_k(\partial\sigma_{ij}/\partial x_l) - \sigma_{ij}(\partial P_k/\partial x_l)\right)/2$. The higher elastic gradient term $\frac{1}{2}v_{ijklmn}(\partial\sigma_{ij}/\partial x_m)(\partial\sigma_{kl}/\partial x_n)$ is necessary for the stability of the thermodynamic potential if the full Lifshitz invariant is included. Application of either the term $F_{ijkl}P_k(\partial\sigma_{ij}/\partial x_l)$ or the term $F_{ijkl}\left(P_k(\partial\sigma_{ij}/\partial x_l) - \sigma_{ij}(\partial P_k/\partial x_l)\right)/2$ results in the same equations of state. The full form, however, leads to the higher order elastic equations and affects the boundary conditions [29, 30, 31, 32, 33].

Polarization distribution can be found from the Euler-Lagrange equations obtained after variation of the free energy (1)

$$a_{ik}P_k + a_{ijkl}P_jP_kP_l + a_{ijklmn}P_jP_kP_lP_mP_n - g_{ijkl}\frac{\partial^2 P_k}{\partial x_j \partial x_l} - Q_{ijkl}\sigma_{kl}P_j + F_{ijkl}\frac{\partial\sigma_{kl}}{\partial x_j} = E_i, \qquad (2)$$

along with the boundary conditions on the top surface of the film $S$ at $x_3 = h$, $\left.\left(g_{kjim}n_k\frac{\partial P_m}{\partial x_j} + a_{ij}^S P_j - F_{jkim}\sigma_{jk}n_m\right)\right|_{x_3=h} = 0$. The direct impact of the flexocoupling is the inhomogeneous terms in the boundary conditions.

Elastic strains are defined by equations, $u_{ij} = -\delta G_V / \delta\sigma_{ij}$. They are expressed via mechanical displacement components, $U_i$, as $u_{ij} = (\partial U_i/\partial x_j + \partial U_j/\partial x_i)/2$: Elastic stress tensor satisfies the mechanical equilibrium equation $\partial\sigma_{ij}/\partial x_j = 0$; The boundary conditions at the mechanically free surface $x_3 = h$ can be obtained from the variation of the free energy (1) with respect to the stresses $\sigma_{ij} n_j\big|_S = 0$. Here $n_j$ are components of the external normal to the film surface. Misfit strain $u_m$ existing at the film-substrate interface ($x_3 = 0$) leads to the boundary conditions for $U_i$

$$\left(U_1 - x_1 u_m\right)\Big|_{x_3=0} = 0, \quad \left(U_2 - x_2 u_m\right)\Big|_{x_3=0} = 0, \quad U_3\big|_{x_3=0} = 0. \tag{3}$$

The periodic conditions were imposed at the lateral sides, $U_1\big|_{x_1=-w/2} - U_1\big|_{x_1=w/2} = w u_m$, $U_2\big|_{x_2=-w/2} - U_2\big|_{x_2=w/2} = w u_m$, while the period $w$ should be defined self-consistently.

The electric field $\mathbf{E}$ is determined self-consistently from the electrostatic problem for the electric potential φ,

$$\varepsilon_0 \varepsilon_b \frac{\partial^2 \varphi}{\partial x_i \partial x_i} = -\frac{\partial P_j}{\partial x_j}, \tag{4}$$

supplemented by the condition of potential continuity at the top surface of the film, $\left(\varphi^{(e)} - \varphi^{(i)}\right)\Big|_{x_3=h} = 0$. The difference of electric displacement components $D_n^{(i)} - D_n^{(e)}$ is conditioned by the surface screening produced by the ambient free charges at the film surface , $\left(D_3^{(i)} - D_3^{(e)} + \varepsilon_0 \dfrac{\varphi}{\lambda}\right)\Bigg|_{x_3=h} = 0$. Here electric displacement $\mathbf{D} = \varepsilon_0 \varepsilon_b \mathbf{E} + \mathbf{P}$, the subscript "$i$" means the physical quantity inside the film, "$e$" – outside the film. The conditions of zero potentials were imposed at the bottom electrode ($x_3 = 0$) and a remote top electrode ($x_3 = H + h$, $H \to \infty$), respectively [34] [see **Fig. 7(a)**].

We suppose that most of defects are located in a thin top layer of thickness $h_0$ beyond which their concentration decreases exponentially towards the film bulk [35] (see **Fig. 7(b)** and Ref. [22]):

$$\delta N(x_3) \approx \frac{N_0}{1 + \exp\left[-\left(x_3 - h + h_0\right)/\Delta h\right]}. \tag{5}$$

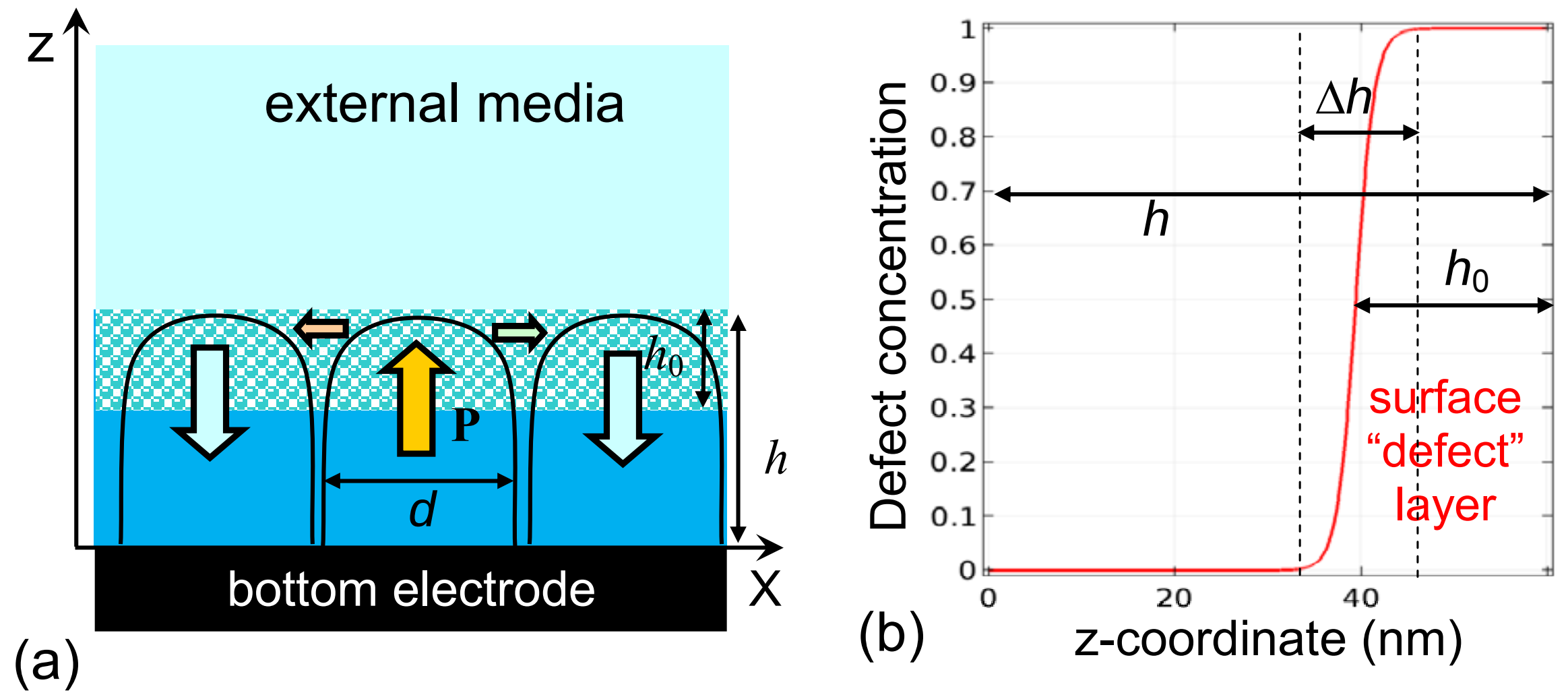

**Fig. 7. (a)** Scheme of a film with the thickness $h$ and the layer of thickness $h_0$ where defects are accumulated. **(b)** Normalized concentration of defects inside the layer of thickness $h_0$ and transition layer depth $\Delta h$. Reprinted from Ref.[22].

### 2.2. Results and discussion

**Figure 8** shows the distribution of polarization components in a ferroelectric $PbTiO_3$ film with a flexo-chemical effect (b, d) and without it (a, c). The absence of the x-component inside the film indicates that the domains are stretched regions from the base of the film to its surface. One component of the polarization smoothly passes into the other near the surface. The distribution of the y-components differs from the model with and without the flexo-chemical coupling. Without the flexo-chemical coupling, the distribution of the y-component is a parallel region. However, the presence of a flexo-effect leads to a distortion of the domain structure. This is because the system, attempts to reduce energy by distortion of the domain structure near the lower electrode due to the flexoelectric effect. In addition, this form of domains creates additional mechanical and electrical stresses that tend to reduce the energy of the system.

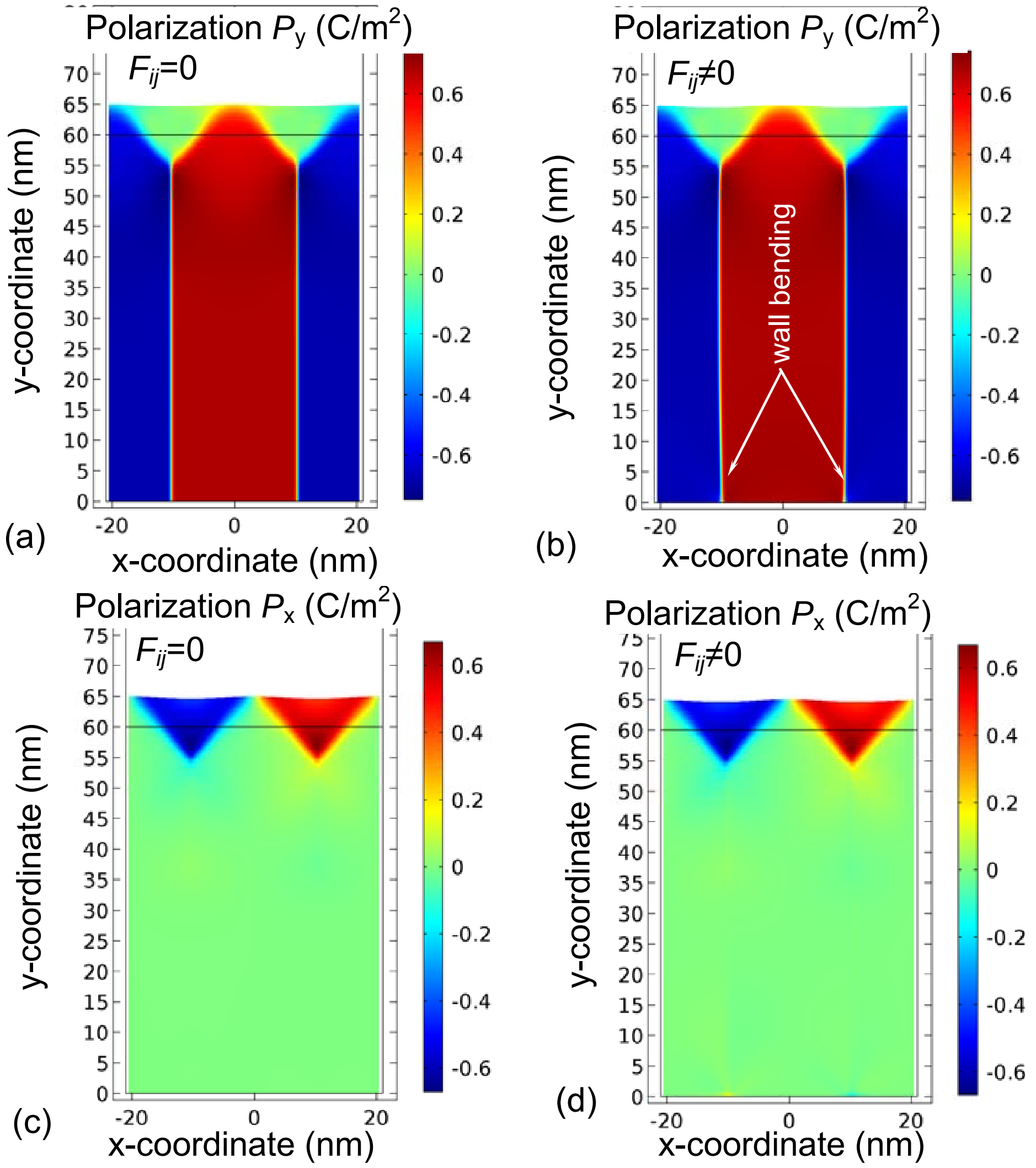

**Fig. 8**. Maps of spatial distribution of electric field components $P_y$ **(a, b)** and $P_x$ **(c, d)** in the cross-section of the ferroelectric $PbTiO_3$ film at room temperature in the cases when flexo-chemical coupling is not included **(a, c)** and included **(b, d)** into consideration. Film thickness $h$= 60 nm and $h_0$=25 nm [see **Fig. 7(b)**]. $N_0 = 2\times10^{26}$ m$^{-3}$, other material parameters are listed in Ref.[22].

**Figure 9** illustrates the distribution of the electric field components at room temperature after a numerical experiment with a flexo-chemical coupling (b, d) and without it (a, c). The distribution of the components exactly repeats the form of the domains in **Fig. 8**. The absence of fields in the middle of the film indicates that the system is in equilibrium and the domain structure encircles the uncharged domain walls, which are parallel or antiparallel to the direction of polarization. Domain walls are not charged near the lower electrode in Fig. 9 (a) (i.e. without the flexo-chemical coupling), but not in **Fig. 9 (b)**. **Figure 3 (b)** shows that the field distribution repeats the distribution of the y-component of the polarization, which indicates the presence of electric fields that try to align the domain walls in the system. The presence of a small field near the contact of the film with the lower electrode is associated with the screening charge at the surface of the material. The distribution and magnitude of the field components are determined by the distribution and magnitude of the polarization components. The asymmetry of the y-component related to compensation processes by the screening charge to reduce the energy of the system.

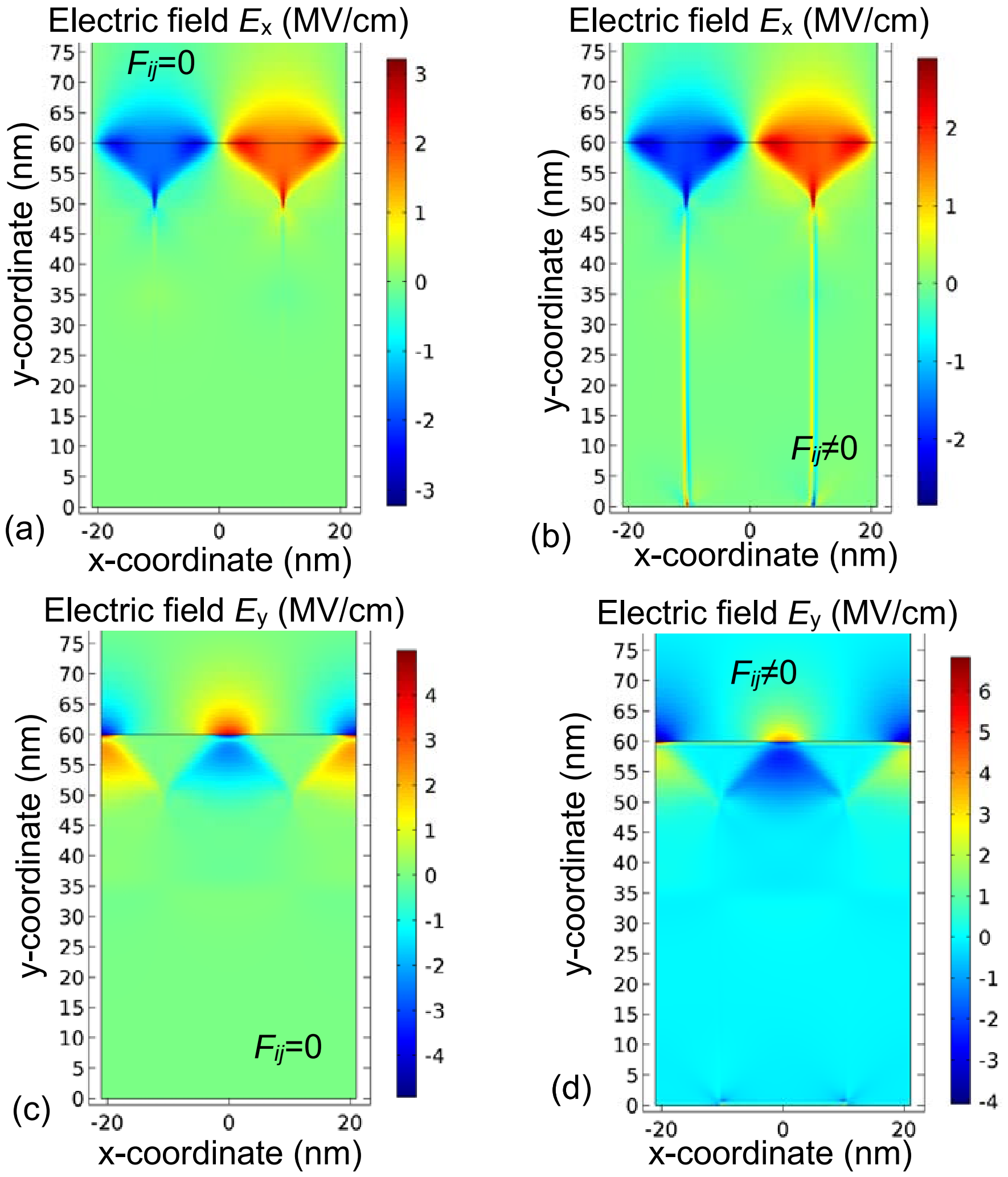

**Fig. 9**. Maps of spatial distribution of electric field components $E_x$ **(a, b)** and $E_y$ **(c, d)** in the cross-section of the $PbTiO_3$ film at room temperature in the cases when flexoelectric coupling is not included **(a, c)** and included **(b, d)** into consideration. Film thickness h= 60 nm and $h_0$=25 nm [see **Fig. 1(b)**]. $N_0 = 2\times10^{26}$ $m^{-3}$, other material parameters are listed in Ref.[22].

**Figure 10** illustrates the distribution of mechanical stress components at room temperature, calculated with the flexo-chemical coupling (b, d) and without it (a, c). The distribution of stress components for plots **10(a)** and 10(**b**) is almost identical. The presence of a negative area near the surface of the film indicates that the system try to reduce its energy due to the compressive stress. The presence of defects leads to decrease the total stress in the surface defect layer. This result explained by the fact that each defect creates a stress of random direction that on average compensates each other. In addition, the main influence has the size and shape of the domain structure. The presence of domain walls lead to increase the energy of the system, but it significantly reduces the energy to the depolarization field, and the influence of the flexo-effect reduces the total energy due to the distortion of the domain shape. This particularly clearly seen near the surface and the contact of the sample with the lower electrode, since these regions are most easily deformed.

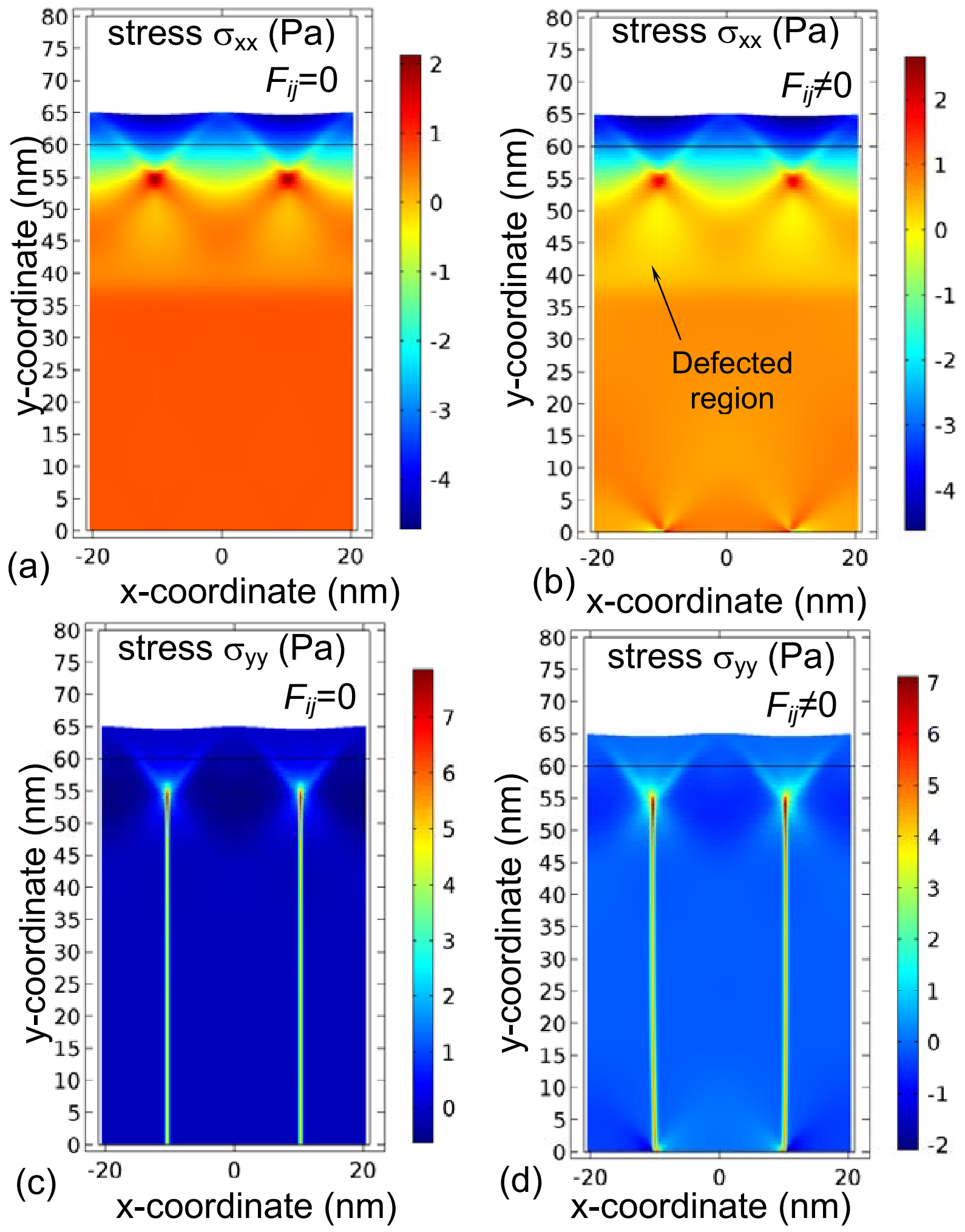

**Fig. 10**. Maps of spatial distribution of elastic stress components $\sigma_{xx}$ **(a, b)** and $\sigma_{yy}$ **(c, d)** in the cross-section of the $PbTiO_3$ film at room temperature in the cases when flexoelectric coupling is not included (a, c) and included (b, d) into consideration. $N_0 = 2\times10^{26}$ m$^{-3}$, other material parameters are listed in Ref.[22].

**Figure 11** illustrates the equilibrium distribution of the elastic strain components at room temperature, calculated with the flexo-chemical coupling (b, d) and without it (a, c). It can be seen on the figures that the strain components completely repeat the shape of the domains, which indicate a distortion of the shape of the sample at the presence of the domain structure. The presence of the flexo-chemical coupling distorts the domain structure, which leads to additional deformations. Thus, the presence of a flexo-chemical couplingleads to the appearance of strains that tend to reduce the energy of the system, that seen from the distribution of field components, stresses, and elastic strains.

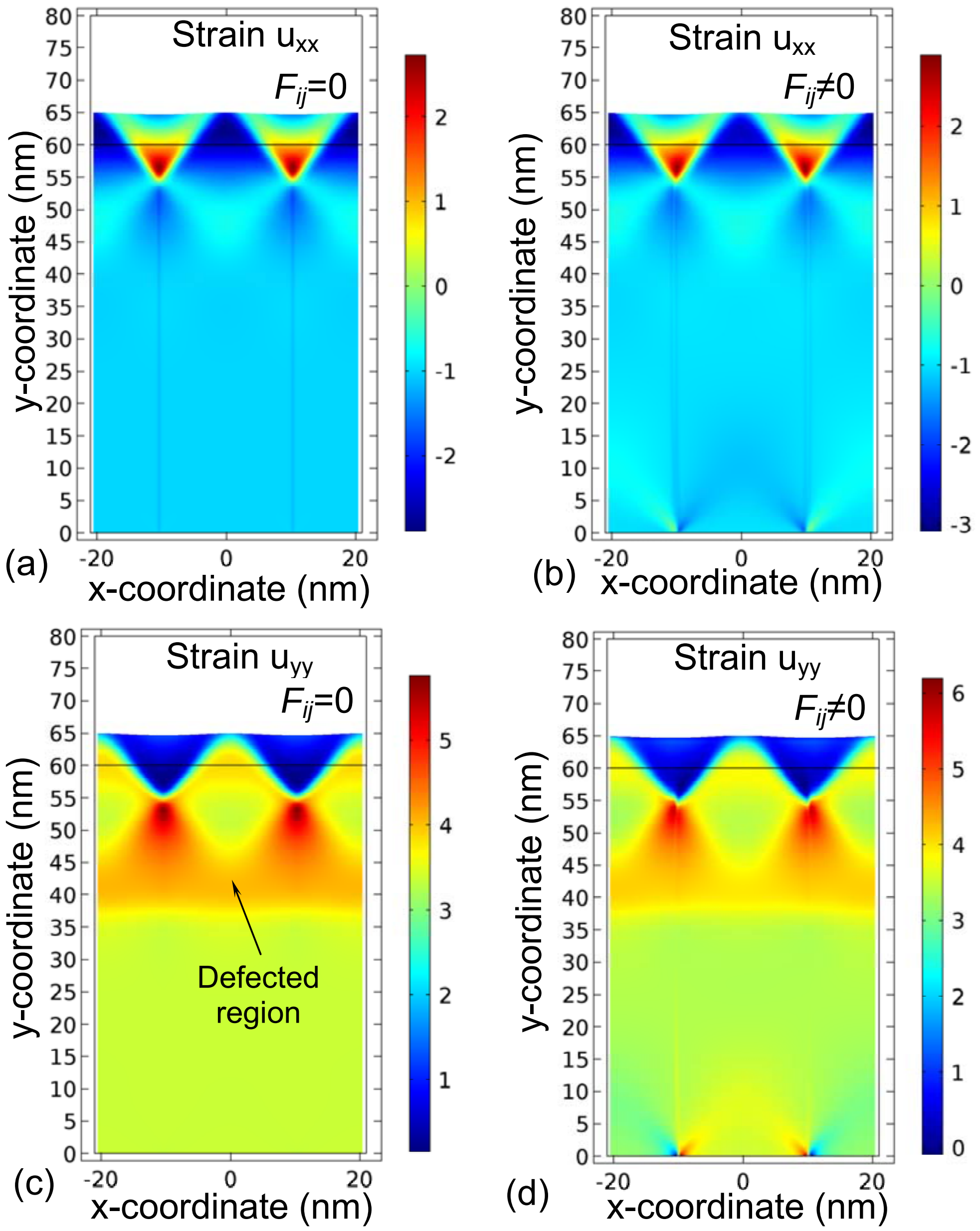

**Fig. 11**. Maps of spatial distribution of elastic strain components $u_{xx}$ **(a, b)** and $u_{yy}$ **(c, d)** in the cross-section of the $PbTiO_3$ film at room temperature in the cases when flexoelectric coupling is not included (a, c) and included (b, d) into consideration. $N_0 = 2\times10^{26}$ $m^{-3}$, other material parameters are listed in Ref.[22].

Using distributions of physical fields, one could calculate the energy of the system as a function of the distance between the domain (see **Fig.12**). It is seen that there is a minimum of energy at certain value of the separation between the domains.

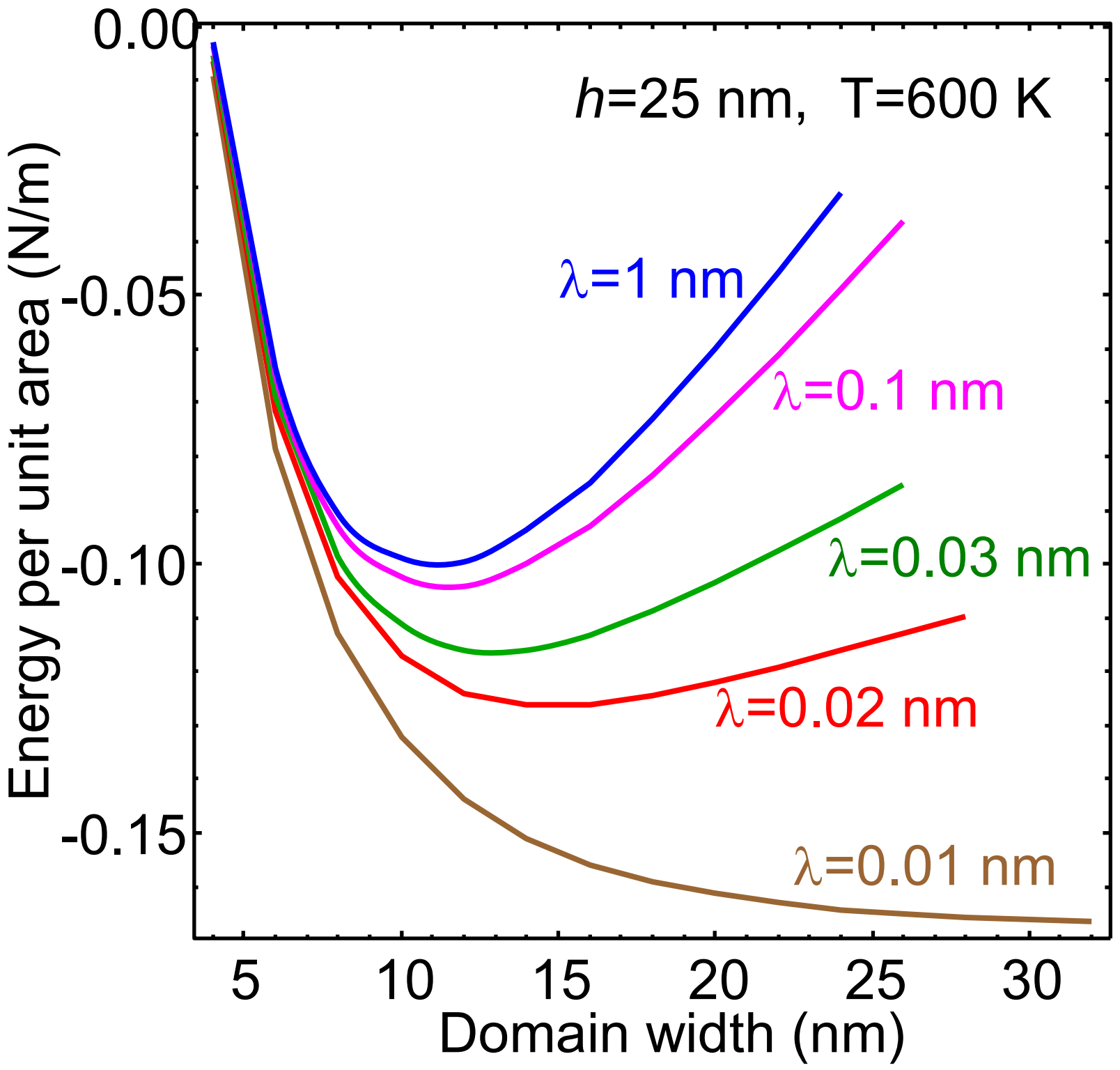

**Figure 12**. The dependence of the total energy on the domain width (potential relief of wall-wall interaction) for different values of surface screening length λ (shown near the curves). Parameters: film thickness is 25 nm at room temperature for $PbTiO_3$ ferroelectric. $N_0 = 2\times10^{26}$ $m^{-3}$, other material parameters are listed in Ref.[22].

The minima of the potential relief determine the equilibrium distance between the domain walls (and hence the domain structure period). Note, that at sufficiently high screening length (absence of screening) the equilibrium period is almost independent on the screening length, while at small λ (almost ideal screening) the equilibrium period rapidly increases, and for the screening length equal to some critical value $\lambda_{cr}$ the equilibrium period diverges so that system transforms into single domain state (there is no equilibrium distance between the domain walls for $\lambda<\lambda_{cr}$).

It is seen, that at small width values wall-wall interaction is almost independent on the screening (short-range interaction due to the correlation energy), while at larger distance between walls electrostatic interaction dominates the interaction potential.

The absence of equilibrium domain structure at sufficiently small λ is explained by behavior of the surface potential with the decrease of λ (see **Fig.13**), namely potential drops to zero independently on the domain width with λ tending to zero. Therefore the driving forces to induce the domain appearance disappear in this limit.

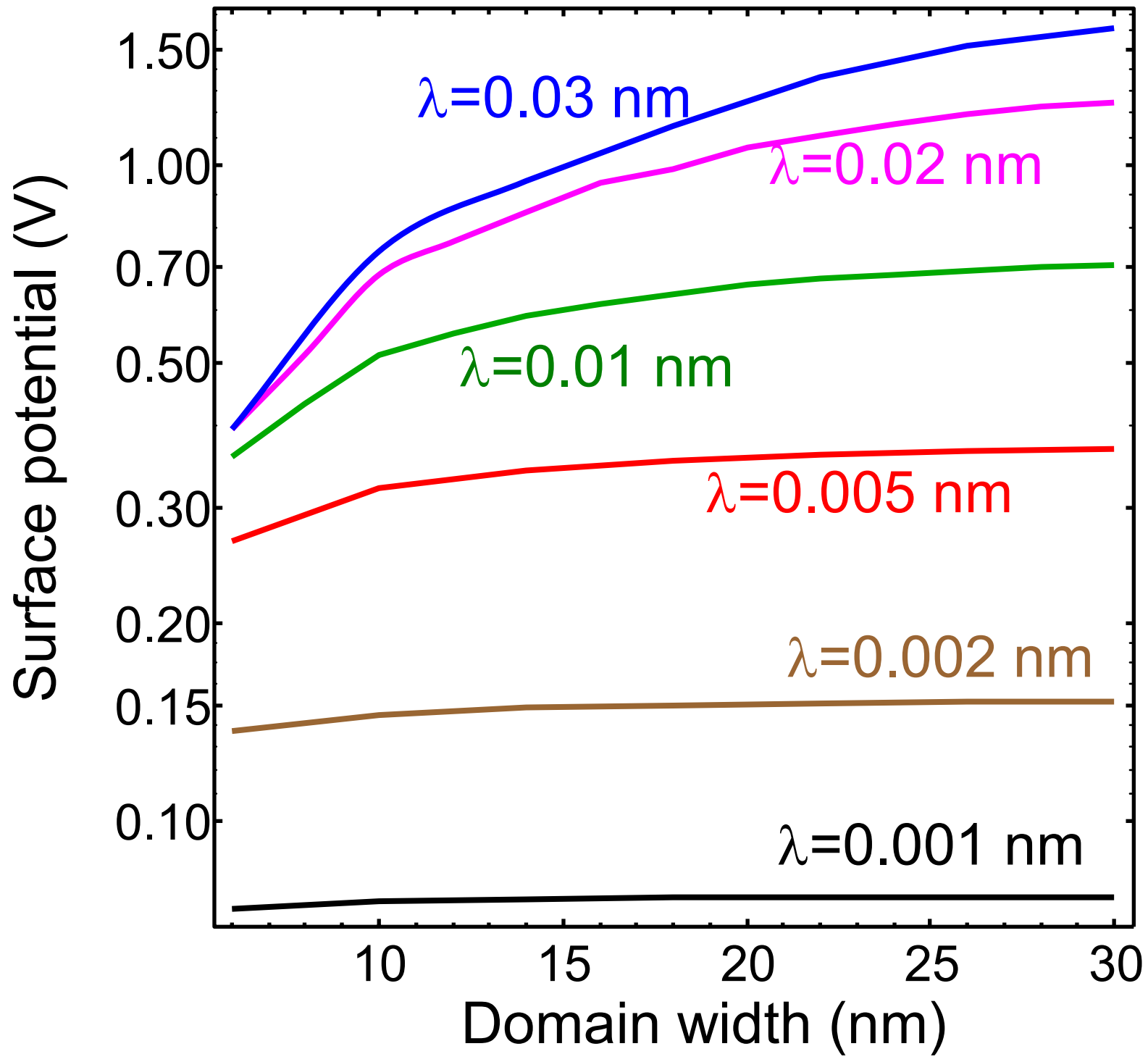

**Fig. 13**. Dependence of surface potential (maximal value) on the domain width for different values of surface screening length λ (shown near the curves). Parameters: film thickness is 25 nm at room temperature for $PbTiO_3$ ferroelectric $N_0 = 2\times10^{26}$ m$^{-3}$, other material parameters are listed in Ref.[22].

**Figure 14** shows that for small values of the length of surface screening, the period of the domain structure is not critical. This can be explained by the fact that for small lengths of screening charge is localized near the surface z = 0 and there is no charge redistribution that is seen on the distribution of the surface potential and field components. However, with increasing of the screening length, the distribution of the surface potential slowly goes out to saturation with increasing the period of the domain structure. This effect indicates that with increasing domain sizes, surface charges tend to reduce the energy of the electric field.

Minimization of the free energy of the system with respect to the domain width allows one to determine the energy of equilibrium state as a function of film thickness, temperature etc.. **Figure 15** illustrates the dependence of the energy of a unit cell on temperature**.** It can be seen that for small temperatures and non-zero defect concentration the energy is slightly smaller indicating the possibility of creating more stable structures by including defects. However, as the temperature increases, the curves increase monotonically and, starting from 480 K, the defects increase the energy of the cell as compared with the defect-free structure. In addition, the figure shows that defects lead to the shift of the phase transition temperature to higher temperatures.

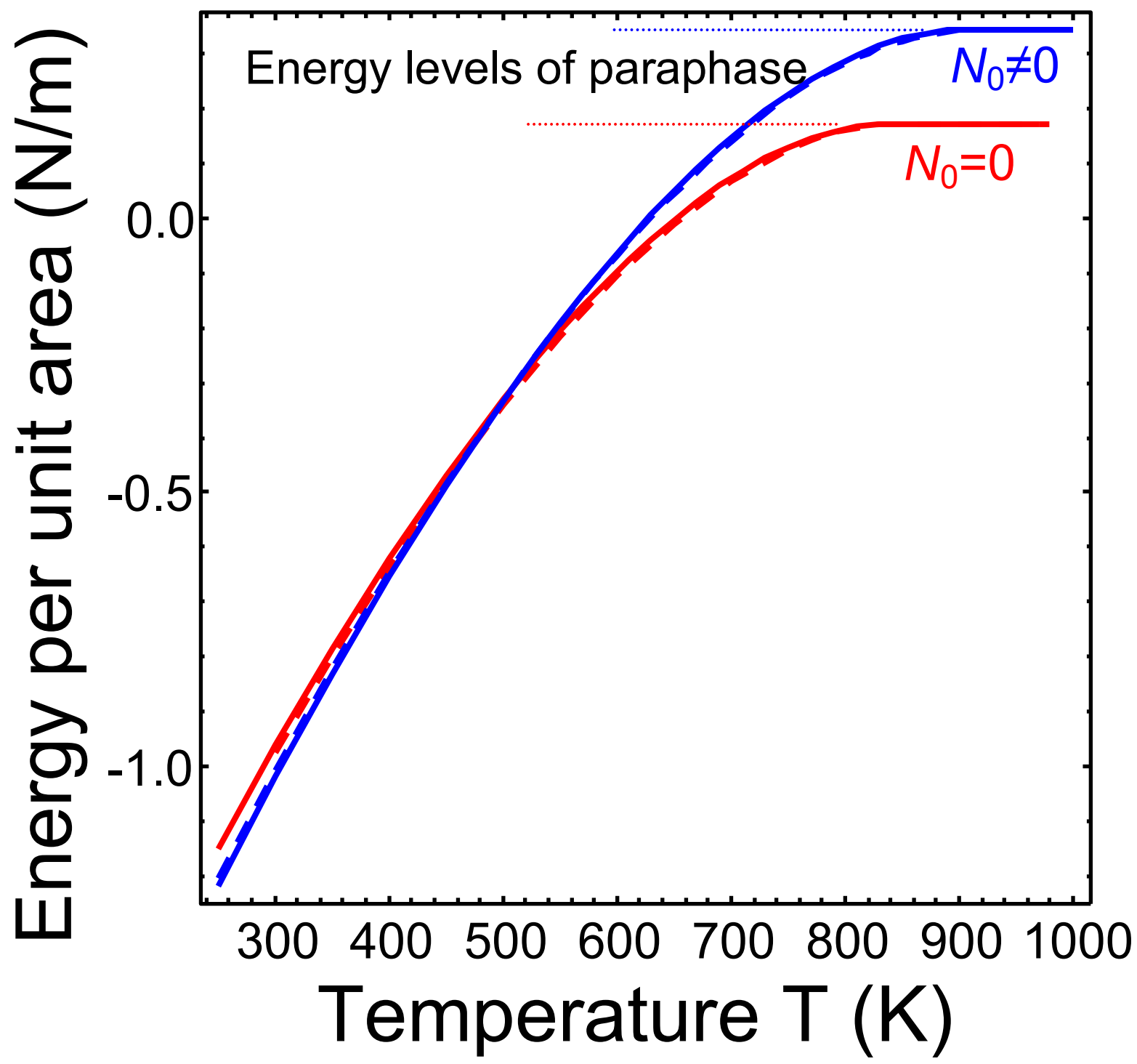

**Fig. 14.** Temperature dependence of total energy of the system for different values of defects concentration. Here two case were considered, when flexoelectric coupling is not included (dashed curves) and included (solid curves) into consideration. Parameters: film thickness is 25 nm at room temperature for $PbTiO_3$ ferroelectric. Screening length λ=0.1nm, $N_0 = 2\times10^{26}$ m$^{-3}$.

**Figure 15** shows the polarization versus temperature, calculated for different film thicknesses. It can be seen that the temperature of the phase transition decreases with decreasing film thickness. However, the presence of defects shifts the Curie temperature to higher temperatures, which makes it possible to control the temperature of the phase transition. In addition, with a further increasing of the film thickness, the temperature of the phase transition does not change and the curves have the same behavior.

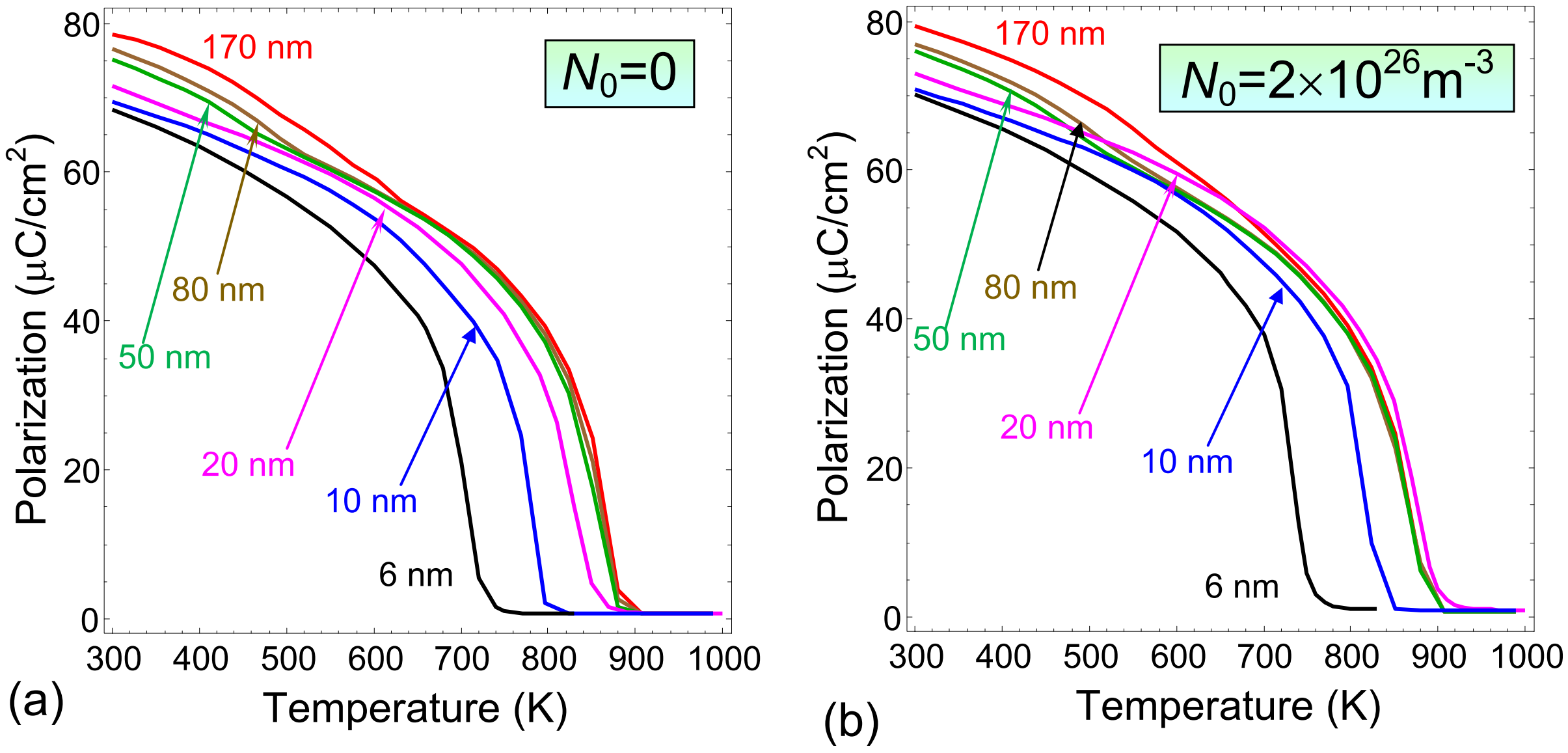

**Figure 15.** Temperature dependence of the maximal spontaneous polarization calculated for different film thicknesses $h$ = (6, 10, 20, 50, 80, 170) nm (labels near the curves) without defects [$N_0 = 0$, plot **(a)**] and for $N_0 = 2\times10^{26}$ m$^{-3}$ [plot **(b)**], screening length λ=0.1nm and the depth of defect layer $h_0$=25 nm. The breaks in the films thicker than 50 nm indicate the appearance of the closure domains at low temperatures.

## III. APPLICATIONS OF FERROICS AND MULTIFERROICS

Ferroelectrics are widely used in nonlinear optics as photorefractive materials for spontaneous and induced birefringence, second harmonic generators, optical phase matching, optical closures, modulators, deflectors and waveguide systems. In optoelectronics and sensorics, ferroics are used as photovoltaic convertors, pyroelectric sensors and irradiation detectors (from gamma irradiation through ultraviolet to infrared) frequency. In information technology, ferroics and multiferroics are indispensable magneto- and electrically-controlled elements of nonvolatile memory; field effect transistors with ferroelectric gate. In microelectronics, high-frequency variconds are variable-capacitance electro-controlled capacitors. End converters are micro-nanomixers and actuators based on piezoelectric effect that are used in microelectromechanics.

In recent years, the interest of scientists and engineers has significantly increased to ferroic relaxors, which have been successfully used in ultrasonic diagnostics and piezo-actuators. This led to the development of methods for controlling properties and advanced improvement of relaxor ferroics performances; expanded areas of their applications, especially as efficient energy storage devices.

Among advanced applications of multiferroics, the special attention is paid to electronic devices that use multiferroics with a large magnetoelectric effect. Due to it, the sensitivity of devices to very small electric and magnetic fields, including human bio-fields, greatly accentuates the sensitivity of existing devices created on the basis of the Hall effect and giant magnetoresistance, in addition, instruments using the magnetoelectric effect are cheaper than others. Today such devices as magnetocardiographs and magnetoencephalographs use multiferroics with the magnetoelectric effect are in the market [8]. Recently discovered reentrant phase, which allows to conserve the useful properties in ultra-small nanoferoics, and hence opened the way to create of multi-layer ceramic capacitors of ultra-high capacity with small size and weight. Magnets, magnetotransport and spintronics (which description goes beyond this review) are of particular importance for applications.

Some perspectives of ferroics and multiferroics applications appear in innovative nanotechnologies, as the newest elements of non-volatile memories based on the huge magnetic-electric coupling in multiferroics, field-effect transistors (FET) with antiferroelectric gate; FET based on graphene-on-ferroelectric. The concept of the "Internet of Things" awaits for new spintronic devices based on ferroelectric semiconductors with the Rashba effect. Nanoactuators can be based on a giant flexo-electrochemical effect in nano-sized paraelectrics. Piezo and pyroelectric nano-generators of electric energy can be based on ordered arrays of ferroelectric nanoparticles can be used for nanoelectromechanics.

## IV. SUMMARY

To conclude ferroics and multiferroics are unique objects for fundamental physical studies of complex nonlinear processes and phenomena which occur in these substances at the micro and nanoscale. Due to the possibility to control physical properties ferroics and multiferroics using size effects, nanoferroics are among the most promising materials for the newest applications in nanoelectronics, nanoelectromechanics, information technologies and nonlinear optics.

## REFERENCES

[1] Nicola A.Spaldin, Manfred Fiebig,. "The renaissance of magnetoelectric multiferroics". *Science*. **309**: 391. (2005)doi:10.1126/science.1113357

[2] R.Ramesh,; Nicola A.Spaldin,. "Multiferroics: progress and prospects in thin films". *Nature Materials.* **6** (1): 21–29. (2007). doi:10.1038/nmat1805

[3] K. Aizu, Possible species of "ferroelastic" crystals and of simultaneously ferroelectric and ferroelastic crystals. *Journal of the Physical society of Japan*, *27*(2), 387-396. (1969). https://doi.org/10.1143/JPSJ.27.387

[4] K. Aizu, Possible species of ferromagnetic, ferroelectric, and ferroelastic crystals. *Physical Review B*, *2*(3), 754. (1970) https://doi.org/10.1103/PhysRevB.2.754

[5] A.N. Morozovska. ABOUT THE STATE AND PROSPECTS OF FERROICS PHYSICS IN UKRAINE. *Visnyk NAS of Ukraine*, **2б**, 42, (2018) ISSN 1027-3239. (https://doi.org/10.15407/visn2018.02.042)

[6] Y.-M. Kim, A. Kumar, A. Hatt, A. N. Morozovska, A. Tselev, M. D. Biegalski, I. Ivanov, E. A. Eliseev, S. J. Pennycook, J. M. Rondinelli, S. V. Kalinin, A. Y. Borisevich. Interplay of octahedral tilts and polar order in BiFeO3 films. *Adv. Mater.*, **25**, 2497–2504 (2013) (10.1002/adma.201204584)

[7] A. N. Morozovska, V. V. Khist, M. D. Glinchuk, Venkatraman Gopalan, E. A. Eliseev. Linear antiferrodistortive-antiferromagnetic effect in multiferroics: physical manifestations. Physical Review **B** **92**, 054421 (2015)

[8] A. P., Pyatakov, A. K.Zvezdin, Magnetoelectric and multiferroic media. Physics–Uspekhi, 2012, 55:6, 557–581 (2012)

[9] D. V. Karpinsky, E. A. Eliseev, Fei Xue, M. V. Silibin, A. Franz, M. D. Glinchuk, I. O. Troyanchuk, S. A. Gavrilov, Venkatraman Gopalan, Long-Qing Chen, and A. N. Morozovska. "A Comprehensive Thermodynamic Potential and Phase Diagram for Multiferroic Bismuth Ferrite". npj Computational Materials 3:20 (2017); doi:10.1038/s41524-017-0021-3

[10] M.D. Glinchuk, E.A. Eliseev, A.N. Morozovska, R. Blinc. Giant magnetoelectric effect induced by intrinsic surface stress in ferroic nanorods. Phys. Rev. B 77, № 2, 024106-1-11 (2008).

[11] Anna N. Morozovska and Maya D. Glinchuk. Reentrant phase in nanoferroics induced by the flexoelectric and Vegard effects. *J. Appl. Phys.* 119, 094109 (2016); doi: 10.1063/1.4942859

[12] Eugene A. Eliseev, Ivan. S. Vorotiahin, Yevhen M. Fomichov, Maya D. Glinchuk, Sergei V. Kalinin, Yuri A. Genenko, and Anna N. Morozovska. Defect driven flexo-chemical coupling in thin ferroelectric films. (*Accepted* to Physical Review B)

[13] E.A. Eliseev, A.N. Morozovska, M.D. Glinchuk, and R. Blinc. Spontaneous flexoelectric/flexomagnetic effect in nanoferroics / Phys. Rev. B. **79**, № 16, 165433-1-10, (2009).

[14] P. Zubko, G. Catalan, A.K. Tagantsev. Flexoelectric Effect in Solids. *Annual Review of Materials Research* 43: 387-421. (2013).

[15] Daniel A. Freedman, D. Roundy, and T. A. Arias, "Elastic effects of vacancies in strontium titanate: Short-and long-range strain fields, elastic dipole tensors, and chemical strain." Phys. Rev. B **80**, 064108 (2009).

[16] Jinlong Zhu, Wei Han, Hui Zhang, Zhen Yuan, Xiaohui Wang, Longtu Li, and Changqin Jin. Phase coexistence evolution of nano $BaTiO_3$ as function of particle sizes and temperatures. Journal of Applied Physics **112**, 064110 (2012)

[17] Eugene A. Eliseev, Anna N. Morozovska, Maya D. Glinchuk, and Sergei V. Kalinin. Missed surface waves in non-piezoelectric solids. Physical Review B 96, No.4, 045411 (2017)

[34] Exactly these conditions correspond to the minimum of the Gibbs potential, see §10 in L.D. Landau and E.M. Lifshitz, Theoretical Physics, Electrodynamics of Continuous Media, Vol. VIII, Pergamon, Oxford, 1963.

[35]. B.W. Sheldon, V.B. Shenoy, Space charge induced surface stresses: implications in ceria and other ionic solids, Phys. Rev. Lett. **106**, 216104 (2011).